\documentclass[a4paper, twocolumn]{article}
\usepackage{amssymb}
\usepackage{amsmath}
\usepackage[font=small]{caption}
\usepackage{subcaption}
\usepackage{dsfont}
\usepackage[pdftex]{graphicx}
\usepackage{stmaryrd}
\graphicspath{{./}{./figures/}}
\usepackage{url}
\usepackage{authblk}
\usepackage{hyperref}
\hypersetup{
	colorlinks,
	citecolor=black,
	filecolor=black,
	linkcolor=black,
	urlcolor=black
}
\usepackage[title]{appendix}
\usepackage[table]{xcolor}
\usepackage{placeins}
\usepackage[nodoi]{apacite}

\newcommand{\I}[1]{\mathds{1}_{#1}}
\newcommand{\keywords}[1] {
\small
\textbf{\textit{Keywords---}}#1
}

\DeclareMathOperator*{\sign}{sign}

\DeclareMathOperator*{\argminop}{argmin}
\newcommand\argmin[1]{\argminop_{#1}}
\DeclareMathOperator*{\argmaxop}{argmax}
\newcommand\argmax[1]{\argmaxop_{#1}}
\newcommand{\norm}[1]{\left\lVert#1\right\rVert}
\usepackage{etoolbox}

\AtBeginEnvironment{APACrefURL}{\renewcommand{\url}[1]{}}
\renewcommand{\doiprefix}{doi:~\kern-1pt}

\begin{document}
\title{Adversarial trading}
\author{Alexandre Miot}
\maketitle
\begin{abstract}
Adversarial samples have drawn a lot of attention from the Machine Learning community in the past few years. An adverse sample is an artificial data point coming from an imperceptible modification of a sample point aiming at misleading. Surprisingly, in financial research, little has been done in relation to this topic from a concrete trading point of view. We show that those adversarial samples can be implemented in a trading environment and have a negative impact on certain market participants. This could have far reaching implications for financial markets either from a trading or a regulatory point of view. 
\end{abstract}
\newline
\hspace{10pt}
\keywords{Adversarial Samples - Machine Learning - Algorithmic Trading}

\section{Introduction}
Adversarial samples have received a lot of attention recently in the Machine Learning (ML) community. We focus here on adverse samples in a trading context and on a market toy model. From \cite{Kurakin2018}, an adversarial sample is ``a sample of input data which has been modified very slightly in a way that is intended to cause a machine learning classifier to misclassify it''. While market manipulation is usually based on a handful of well known techniques, see \cite{Putnins2018}, one's could harness the recent progress of ML on adversarial samples to build more elaborate techniques, see \cite{Wiyatno2019} or \cite{Machado2020} for a review. We call adversarial trading the use of adversarial samples to trade. Financial markets are especially exposed due to the high quantity of data and the pervasiveness of algorithmic trading. On a simulated order book, we investigate the possible implications of adversarial trading.
\section{Related Work}
Since the seminal work of \cite{Dalvi2004}, \cite{Lowd2005} and the more recent \cite{Szegedy2013},  adversarial samples is an active field of research in Machine Learning. It has been explored in computer vision \cite{Moosavi2015}, human vision \cite{Elsayed2018}, real 3D objects \cite{Athalye2017}, tabular data \cite{Ballet2019}, time series \cite{Mode2020}, speech \cite{Carlini2018} or from a theoretical point of view in \cite{Dohmatob2019}, \cite{Tsipras2018}, or \cite{Fawzi2017}. See \cite{Wiyatno2019} for a review.
Yet, even if \cite{Faghan2020} has a similar point of view, to our knowledge little has been done to explore the implications of adversarial samples in financial markets from a practical trading perspective.

\section{Framework}
To measure the efficiency of our adversarial samples, we first build a market simulator. We then build a dataset and train a classifier predicting the direction of the next market move. From adverse samples, we finally build an adversarial trading agent and show its impact on other agents.
\subsection{Market and central limit order book}
We build a simple matching engine to simulate a central limit order book. A trading round is composed of two consecutive steps:
\begin{enumerate}
\item no more book orders can be fulfilled. All market participants determine which quotes they want to send and which unmatched quotes they want to cancel
\item they place their quotes and cancels into the book: orders are executed partially or fully as soon as they match.
\end{enumerate}
At the end of the second step, all matching orders have been executed and we are back to the first step. A simulation is a set of trading rounds for a set of agents. In the following, we will note $b^j_t$ and $o^j_t$, the bid and offer levels at  the $j^\text{th}$ depth and trading round $t$. $qb^j_t$ and $qo^j_t$ are the respective quantities. A typical order book will be\footnote{in our simulations average depth is around 17}:
\begin{equation*}
\begin{array}{cc|cc}
\multicolumn{2}{c|}{\text{Bids}} & \multicolumn{2}{c}{\text{Offers}}\\
\text{Level} & \text{Quantity} & \text{Level} & \text{Quantity}\\
b^1_t & qb^1_t & o^1_t & qo^1_t \\
b^2_t & qb^2_t & o^2_t & qo^2_t \\
&  & o^3_t & qo^3_t \\
\end{array}
\end{equation*}
\subsection{Agents}
Market participants are named agents. Four types of agents are defined:
\begin{description}
	\item[investor agents] have predetermined and constant bearish or bullish view called bias i.e. a bullish agent will only buy the market and a bearish one only sell. At each round, they cancel any previous unmatched order and place a buy or sell order according to their bias
	\item[market maker agents] place a buy and a sell order for the same size at each round cancelling any previous unmatched order
	\item[noisy agent] place random buy and sell orders at each round
	\item[adversarial agent] place adverse orders at each round.
\end{description}
In order to have a balanced market, for each simulation, the number of bullish and bearish investor agents is roughly the same. Each agent has a stop loss. If breached then all its unmatched orders are cancelled and it will not be able to trade any more.
\subsection{Price formation}\label{subsection:priceformation}
Before placing orders, a reference price is computed from the market book. This is essentially the mid market price but restricting price jumps:
\begin{align*}
bm_t &= \frac{1}{2} \left( b^1_t + o^1_t \right) \\
\widetilde{bm_t} &= \left\{ \begin{array}{ll}
	bm_t &\text{ if } \left|\frac{bm_t}{\frac{1}{n} \sum_{i=1}^{n} m_{t-i}} - 1\right|< \alpha\\
	m_{t-1} & \text{ otherwise}\\
	\end{array}\right.\\
m_t &= m_{t-1} \min \left(1 + \alpha, \max \left(1-\alpha, \frac{\widetilde{bm_t}}{m_{t-1}}\right) \right)\text{ ,}
\end{align*}
where $m_t$ is the reference price at trading round $t$, $b^1_t$ and $o^1_t$ being the best bid and offer at $t$, $n=10$ and $\alpha=10\%$.
Bearish (respectively bullish) investor agents place a sell (respectively buy) order randomly around this reference price plus (respectively minus) one percent.\\
Market maker agents use the best bid and offer but skew their quotes depending on the market imbalance. If the market has more quantity on the bid (respectively on the offer) side then the market maker will bid (respectively offer) more aggressively. Noting $mb_t$ and $mo_t$ the bid and offer levels of the market maker:
\begin{align*}
i &= \sum_{j=1}^{n} qb^j_t - \sum_{j=1}^{m} qo^j_t \\
\iota & = \I{i> 0} - \I{i< 0} \\
mb_t &= b^1_t \times 0.95 \times (1+ \iota \alpha) \\
mo_t &= o^1_t \times 1.05 \times (1+ \iota \alpha)\text{ .}
\end{align*}
Noisy or adversarial agents might place orders at any depth in the book either randomly for the noisy agent or according to an adversarial strategy for the adversarial agent.
\subsection{Adversarial trading}

\subsubsection{Surrogate model}
Adverse agents place orders which are designed to have an adverse effect on other market participants. In our framework, we know that modifying the market imbalance might impact market maker agents. The adverse agent, though, has no access to the market maker strategy. For all agents, only the anonymous order book is known.
We build a surrogate model: a classifier which will try to predict if the market will move up or down at the next trading round. We use a logistic regression on features $X_t$ being the concatenation of bid levels, bid quantities, offer levels and offer quantities. The order book can have any depth which can vary with time. We choose a constant depth of 20 adding missing values of -1 for prices and 0 for quantities if the book depth is smaller than 20:
\begin{multline*}
X_t = (b^1_t, b^2_t, -1, \ldots, qb^1_t, qb^2_t, 0, \ldots, \\ o^1_t,o^2_t,o^3_t,-1,\ldots, qo^1_t, qo^2_t, qo^3_t, \ldots)\text{ , }
\end{multline*}
as a result $X_t \in \mathbb{R}_{+}^{80}$.
The surrogate model has two important prerequisites:
\begin{enumerate}
	\item an agent must be able to have an influence on the features
	\item the predicted quantity must be a relevant trading feature to the targetted agents.
\end{enumerate}
On test data our surrogate model has a precision around 70\%.

\subsubsection{Adversarial samples}
\begin{figure}[h]
	\centering
	\includegraphics[width=0.9\columnwidth]{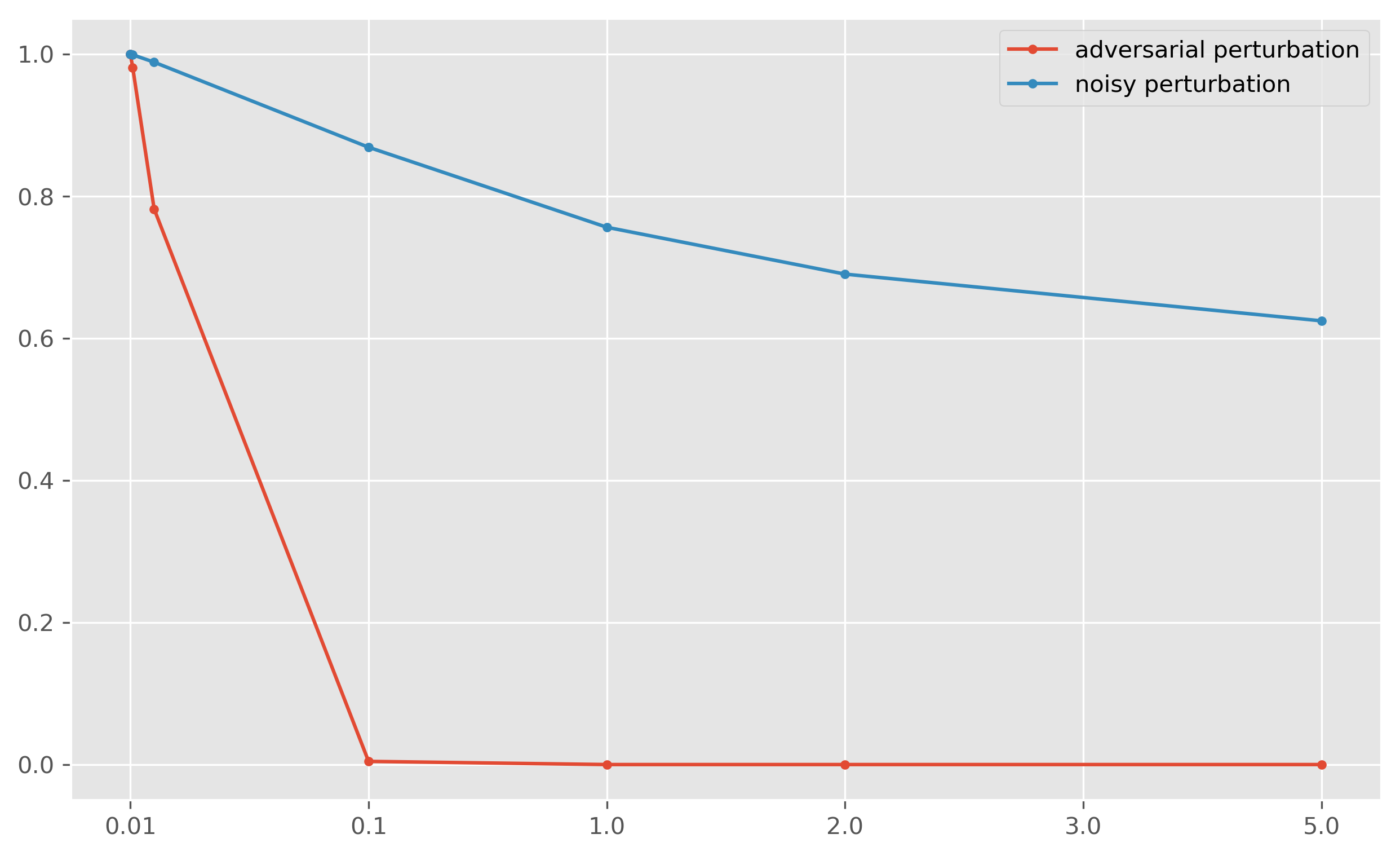}
	\caption{Accuracy loss as the amplitude of the perturbation $\epsilon$ increases. Adversarial perturbations (red) are much more effective than adding noise (blue) of the same amplitude.}
	\label{fig:adversarial_vs_noise}
\end{figure}

We then build adversarial examples using FGSM as described in \cite{Goodfellow2014}. We choose samples correctly predicted by the surrogate model and build for each sample an adverse perturbation defined as
$$
\delta_x = \argmin{f(x+r)\neq f(x)}{||r||}\;\text{,}
$$
and approximated in the FGSM method as 
$$
\delta_x  = \epsilon \, \nabla_r L(x+r, f(x))\text{ ,}
$$
where $f$ is the surrogate classifier, $x$ a correctly classified sample, $\epsilon>0$ a given constraint and $L$ is the classifier cost\footnote{here the negative log-likelihood of the logistic regression}. Supposing that we impose $\norm{r}_{\infty} < \epsilon$, we want to maximize the loss:
$$
\argmax{\norm{r}_{\infty} \leq\epsilon}{L(x+x)-L(r)} \sim \argmax{\norm{r}_{\infty} \leq \epsilon} {r\cdot \nabla L} = \epsilon \sign \nabla L\text{ .}
$$
Our adversarial sample is fast and simple but other constraints might be needed. For example, imagine that the perturbation is negative for the offer quantity at the $j^\text{th}$ depth: it would mean that we should place orders to decrease this quantity. This is impossible in practice without massively impacting the book. One way to handle constraints on adversarial perturbations would be to use techniques described in \cite{Carlini2016}. Another possibility would have been to use Adversarial Transformation Networks of \cite{Baluja2017}. In section \ref{subsubsection:adversarial_agent} we describe how the adversarial agent deals with possibly unrealistic perturbations.\\
It is important to note that noise and adversarial perturbations are very different. On figure \ref{fig:adversarial_vs_noise}, we plot the accuracy of initially correctly predicted samples as we increase the amplitude of perturbation for both noisy and adversarial perturbations. Noise is detrimental to predictions but far less than adversarial samples for a given $\epsilon$ bound. 
\\
In a more general setting, building adversarial samples might be slow. We build a map $x \mapsto \delta_x$ mapping a sample to an approximate perturbation.
We train a multiple target classifier to predict the signs of the gradient $\eta_x = \sign \nabla L_x$. A random forest classifier gives very good results. This classifier is named adversarial estimator. 

\subsubsection{Adversarial agent}\label{subsubsection:adversarial_agent}
The final step is to build an adversarial agent. The task of mapping an adverse perturbation in feature space into actual orders is not obvious. As we have seen, the perturbation might indicate to lower the price at the third depth or decrease the quantity at the fifth depth, for example.
From our adversarial estimator we retrieve a vector $\eta_x\in\{-1, 1\}^{80}$ given by the previously trained multi-target random forest classifier:
$$
\eta_x = (\underbrace{1, 1, \ldots}_{\substack{\text{bid price}\\i=1\ldots 20}}, \underbrace{-1, 1, \ldots}_{\substack{\text{bid quantities}\\i=20\ldots 40}},\underbrace{-1, -1, \ldots}_{\substack{\text{offer prices}\\i=40\ldots 60}},\underbrace{-1, 1, \ldots}_{\substack{\text{offer quantities}\\i=60\ldots 80}})\text{ .}
$$
The adversarial agent discards information given by prices i.e. it discards the values $\eta^i_x$ where $i=1, \ldots ,20$ and $i=40, \ldots, 60$. For $i\in\llbracket 20, 40 \rrbracket$, if the sign of the bid quantity $\eta^i_x$ is +1 then a bid order at this depth level is added for 1 lot (see table \ref{tab:table_mean_sizes} for averages sizes in the book). If the sign is -1 then an offer order at the corresponding depth is placed at the offer price for this depth. The same thing is done for $i\in\llbracket 60, 80 \rrbracket$.
All previous orders from previous trading round are cancelled.
The noisy agent works exactly the same way except that instead of getting $\eta_x\in\{-1, 1\}^{80}$ from the adverse estimator it draws them randomly.
\begin{table}[h]
	\centering
	\footnotesize
	\begin{tabular}{ l | c r }
		& Bid & Offer \\
		\hline
First depth &  $\sim 19$ &  $\sim 19$ \\
Second depth & $\sim 17$&  $\sim 17$ \\
Third to eleventh depth & $\sim 16$ & $\sim 16$ \\
Twelfth to sixteenth depth & $\in [10, 16]$&$\in [10, 16]$ \\
Seventh to twentieth & $\in [4, 10[$  & $\in [4, 10[ $

	\end{tabular}
	\caption{Average book quantities up to 20\textsuperscript{th} depth.}
\label{tab:table_mean_sizes}
\end{table}

\section{Results}
\begin{figure}[h]
	\centering
	\includegraphics[width=0.9\columnwidth]{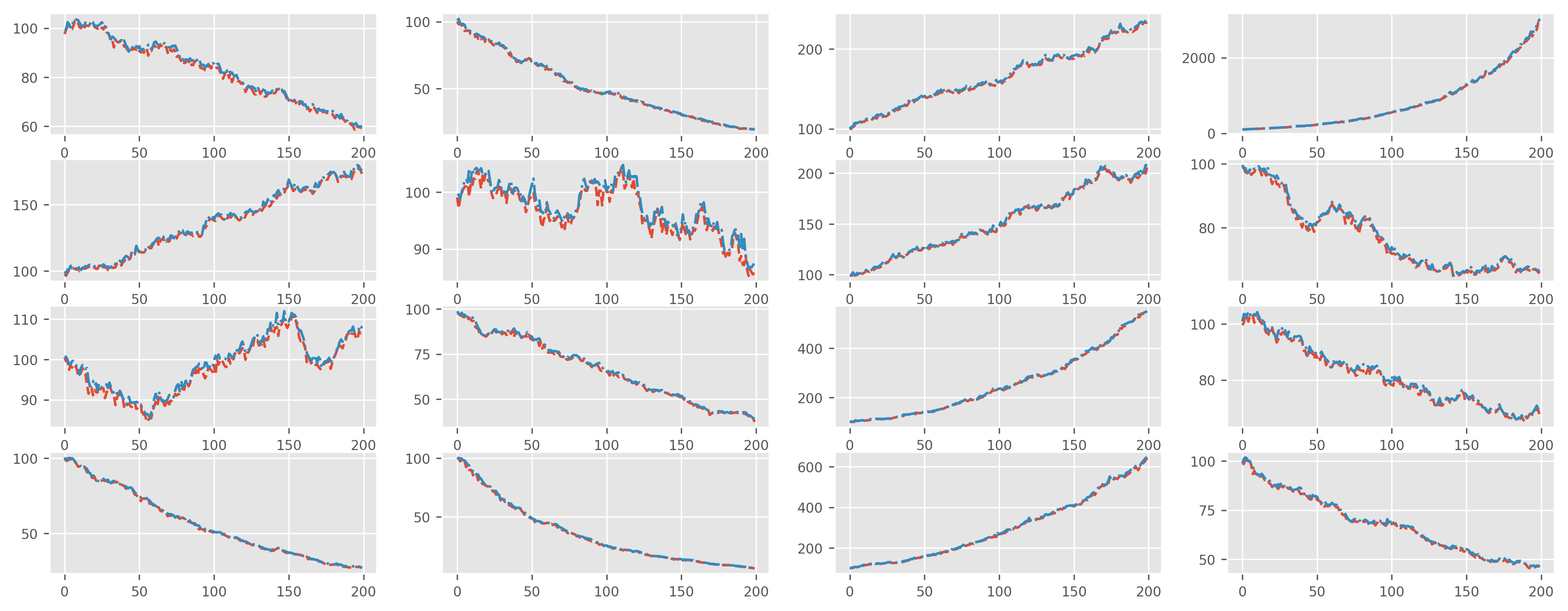}
	\caption{Randomly selected time series of bid (dash red) and offer (point and dash blue) of the first depth of simulated books.}
	\label{fig:trajectories}
\end{figure}

We run 10 rounds of 500 simulations each comprising 200 trading rounds for each of the following setups:
\begin{itemize}
	\item[-] 40 investor agents and one market maker agent
	\item[-] 40 investor agents, one market maker agent and one noisy agent
	\item[-] 40 investor agents, one market maker agent and one adversarial agent.
\end{itemize}
It gives us a total of $10\times 500\times 200=$ 1 million trading rounds for each of the three setups.
Simulated trajectories are not too dissimilar from financial time series as seen in figure \ref{fig:trajectories} and are good enough for our purpose.
The market maker agent takes into account the market imbalance when quoting. As a result, an adversary might be able to impair the market maker profit and loss (P\&L) by placing small orders in the book. This is what we see from our simulations in table \ref{tab:pl_mm_statistics} looking at the cumulative P\&L of the market maker agent under our three setups.
We see that the market maker's performance deteriorates when an adversarial or noisy agent is in the market. Consistently, for all simulations:
\begin{itemize}
	\item[-] the median P\&L of the market maker agent is higher without noisy or adverse agent
	\item[-] the median P\&L of the market maker agent is higher with noisy agent than with adverse agent.
\end{itemize}
Yet, the difference between the noisy and adverse agent setups can be small.
Also, the adversarial agent does not manage to benefit from fooling the market maker as its average P\&L is always negative around -50,000. Actually, the investor agents are benefiting from the adverse agent with an average P\&L going from -4,000 without adverse agent to -2,000 with an adverse agent.
As a result, if one's wanted to build a profitable adversarial agent more work is necessary. Yet, this is not our aim here and we think that these results demonstrate that adversarial trading can work in a concrete trading setup.
\begin{table}[h]
	\centering
	\footnotesize
\begin{tabular}{p{0.2\columnwidth} | p{0.10\columnwidth} p{0.10\columnwidth} p{0.14\columnwidth} p{0.14\columnwidth}}
& Mean & Median & $1^\text{st}$ quartile & $3^\text{rd}$ quartile \\
\hline
Without adversary& 177,000 & 54,000 & 39,000 & 199,000 \\
With noisy agent& 71\,\% & 95\,\% & 80\,\% & 70\,\% \\
With adversary agent&  67\,\% & 92\,\% & 72\,\% &  70\,\% \\
\end{tabular}
	\caption{Rounded P\&L statistics of the market maker agent under different setups (in percentage of P\&L without adversary for adversarial setups).}
\label{tab:pl_mm_statistics}
\end{table}

\section{Discussion}
We have seen that on a simulated market, we can design adversarial agents which actions, though difficult to notice, can have a negative impact on other agents. A legitimate question is to ask if this can be transposed into real markets. To answer this question two distinct aspects have to be taken into consideration:
\begin{enumerate}
	\item the ability to find adversarial samples
	\item the ability to find adversarial agents able to ``implement'' these adverse samples.
\end{enumerate}
On the first point, even if research is still very active, several theoretical articles point to a similar direction involving that it is probably easier in real conditions. The most relevant theoretical work to our topic is \cite{Fawzi2018}: any classifier on a feature space which can be approximated by a highly dimensional generative model is prone to adverse attacks. Latest progress in market generators, see \cite{Wiese2020}, \cite{Koshiyama2020} or \cite{Buhler2020} tend to show that market prices can be approximated by generative models. Moreover, a more complex classifier distinguishing more market conditions is even less robust, the robustness decreasing with the number $K$ of classes.
Also, under mild assumptions, transferability of adversarial samples is granted blurring the distinction between white box and black box attacks. This entails that training a surrogate model might have high probability of working if the attacked model has low generalization error\footnote{which is a precondition to have an algorithmic model in production in the first place}.

On the second point, to our knowledge little research has been carried out. We think though that our findings show that they might exist in practice. As a result, regulators might want to tackle the subject, knowing that in complex markets tracking these adversarial agents might well be difficult, see \cite{Wang2020} though.
 


\section{Conclusion}
In this article, we found that in a simulated framework it is possible to use adversarial samples and implement them as a trading strategy in order to negatively impact some market participants. While finding adversarial samples is a relatively easy task, implementing them in a trading environment is more difficult but as we have seen possible. Importantly, regulators might want to scrutinize application of adversarial trading to insure market integrity. Finally, one could ask if adversarial samples do apply to human brain too. Empirically, it does not seem to be the case and it might also be true in a trading environment: human traders might be less efficient than machines but way sturdier.
\section{Declaration of Interest}

The author report no conflicts of interest. The author alone is responsible for the content and writing of the paper.
\clearpage

\bibliographystyle{apacite}
\bibliography{adversarial_trading}

\end{document}